\documentclass[pre,preprint,showpacs,superscriptaddress]{revtex4-1}
\usepackage{graphicx}
\usepackage{amssymb}
\usepackage{amsmath}
\usepackage{wasysym}
\usepackage{color}
\begin{document}
\title{Sex-ratio bias induced by mutation}
\author{Minjae Kim}
\affiliation{Department of Physics, Pukyong National University, Busan 48513,
Korea}
\author{Hyeong-Chai Jeong}
\affiliation{Department of Physics and Astronomy, Sejong University, Seoul
05006, Korea}
\author{Seung Ki Baek}
\email[]{seungki@pknu.ac.kr}
\affiliation{Department of Physics, Pukyong National University, Busan 48513,
Korea}

\begin{abstract}
A question in evolutionary biology is why the number of males is approximately
equal to that of females in many species, and Fisher's theory of equal
investment answers that it is the evolutionarily stable state.
The Fisherian mechanism can be given a concrete form by a genetic model
based on the following assumptions: (1)
Males and females mate at random. (2) An allele acts on the father to determine
the expected progeny sex ratio. (3) The offspring inherits the allele from
either side of
the parents with equal probability. The model is known to achieve the 1:1 sex
ratio due to the invasion of mutant alleles with different progeny sex ratios.
In this study, however, we argue that mutation plays a more subtle role in that
fluctuations caused by mutation renormalize the sex ratio and thereby keep it
away from $1:1$ in general. This finding shows how the sex ratio is affected by
mutation in a systematic way, whereby the effective mutation rate can be
estimated from an observed sex ratio.
\end{abstract}


\maketitle

\section{Introduction}

The number of males per female is close to one in the world population, and the
value has been found stable across many countries~\cite{james1987human}. This
1:1 sex ratio at birth
is also commonly observed in many other sexually reproducing species.
This is indeed highly nontrivial in that the ratio is suboptimal from the
viewpoint of the population: As far as the growth rate is concerned, which is
directly related to reproductive success of the species, it would
be more efficient
to produce more females than males because females can give birth to offspring.
This female-biased
state cannot be sustained, however, and the reason can be understood from
the ``selfish-gene'' point of view. Along this line, Fisher's theory states that
the one-to-one ratio between males and females is the evolutionarily stable
state in this game of genes~\cite{fisher1930genetical}.
The argument goes as follows~\cite{shaw1953selective}: Consider an individual
with $n$ offspring, of which $nx$ are male and the others are female ($0 \le x
\le 1$). This
individual's next generation has $K$ offspring in total, where $KX$ and $K(1-X)$
are the numbers of males and females, respectively ($0< X <1$). In this case,
the relative investment of the individual is $C_\text{inv} = n/(2K)$ because
we assume that an offspring
inherits one half of the genes from either parent. The focal individual's
genetic contribution to the population is $C_1 = C_\text{inv}$,
which is a reference point to judge an individual's genetic success.
The situation becomes different in the second next generation:
If males and females mate randomly, then the focal individual's
genetic contribution is calculated as
\begin{equation}
C_2 = \frac{1}{4} \left( \frac{nx}{KX} + \frac{n \tilde{x}}{K \tilde{X}}
\right),
\label{eq:shaw}
\end{equation}
where $\tilde{x} \equiv 1-x$ and $\tilde{X} \equiv 1-X$.
According to this formula,
if $X$ exceeds $1/2$, then $C_2$ is greater than $C_\text{inv}$ for $x < X$. By
symmetry, it is also obvious that $C_2 > C_\text{inv}$ for $x > X$ if $X$ is
less than $1/2$. It thus follows that it is genetically beneficial to ``invest''
in the rare sex, which constitutes the basic mechanism for maintaining the
Fisherian sex ratio of 1:1. In this sense, the sex-ratio problem is an example
of conflict between individual and collective
interests~\cite{rankin2007tragedy}.

The Fisherian mechanism has many subtleties, and still not
much is known about deviations from its
prediction~\cite{james2008evidence,west2009sex}. In particular, it is
noteworthy that if the population achieves this predicted ratio, i.e., $X=1/2$,
$C_2$ of Eq.~\eqref{eq:shaw} becomes $C_\text{inv}$ regardless of an
individual's $x$ as long as
the population size is large
enough~\cite{kolman1960mechanism,*verner1965selection,*taylor1980selective}.
It implies that the timescale of this evolutionary dynamics may actually
diverge as the restoring force toward $X=1/2$ vanishes at this point.
Put differently, if mutation occurs with a rate $\mu$, the timescale
would be of an order of $\mu^{-1}$, and Fisher's ratio $X=1/2$ can be achieved
in a limit of $\mu \to 0$. If $\mu$ is small yet finite, on the other hand, the
1:1 ratio may not be reached within finite time.

In this work, we show that a
dynamic equilibrium out of 1:1 actually forms in a minimal model
devised for Fisher's theory. This is counterintuitive
because mutation is an essential ingredient of the Fisherian mechanism.
In short, it cannot work without mutation, and it cannot work with it either.
We will explain this observation in the following way: In the next section, we
introduce a genetic model and study it with three different approaches:
Monte Carlo simulation, integrodifference equations, and renormalization
analysis. We discuss the implications in
Sec.~\ref{sec:discuss} and then conclude this work in Sec.~\ref{sec:summary}.

\section{Genetic model}

\subsection{Monte Carlo simulation}

\begin{figure}
\includegraphics[width=0.49\columnwidth]{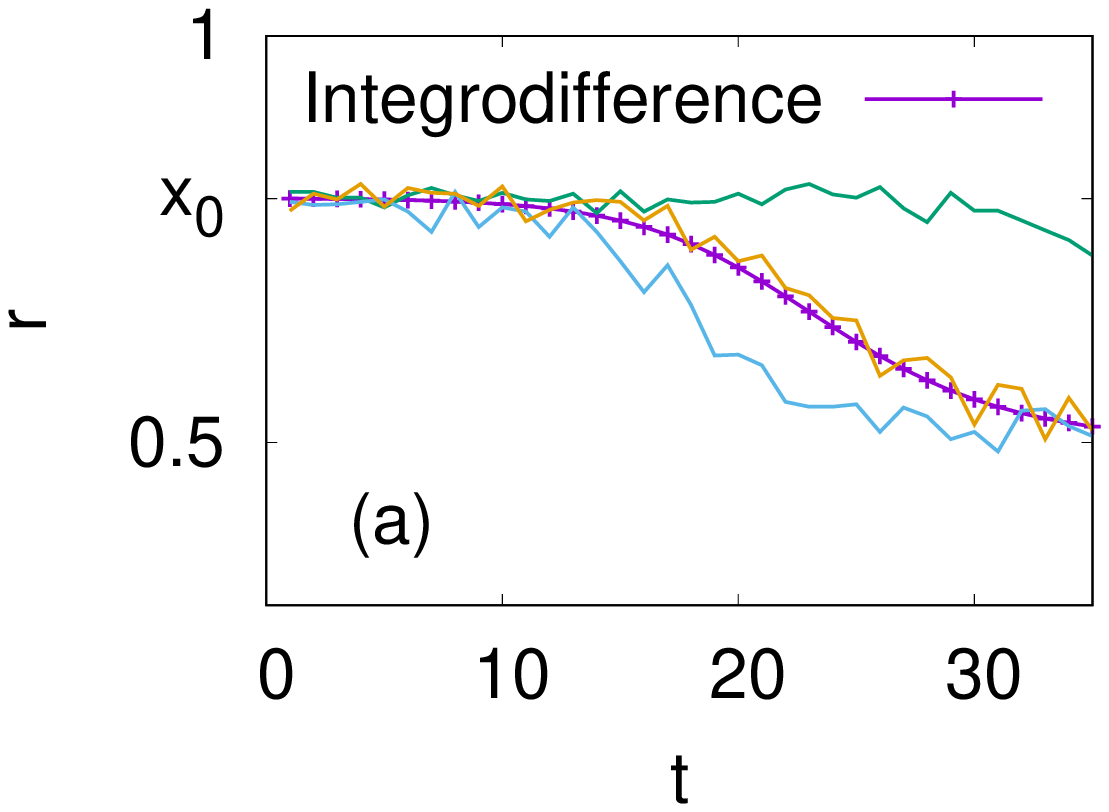}
\includegraphics[width=0.49\columnwidth]{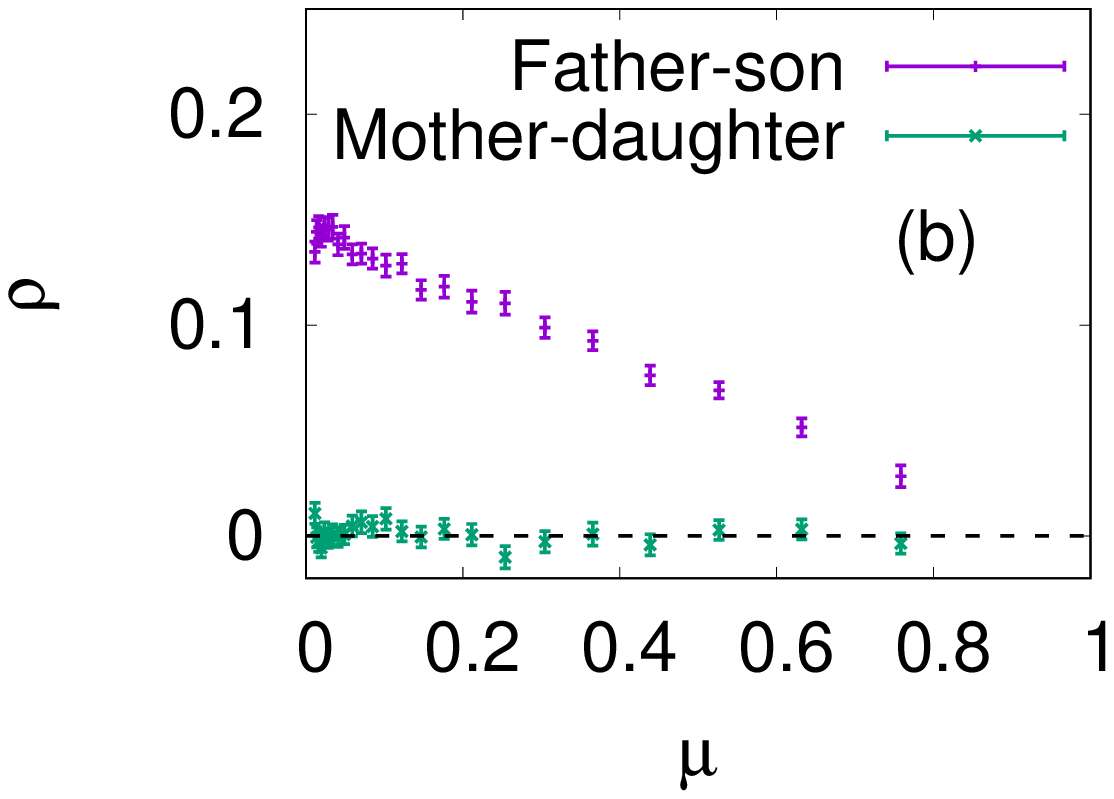}
\caption{(Color online)
(a) Time evolution of the male fraction with $\mu =
10^{-3}$. The horizontal axis represents time in units of generations.
Initially, every individual has an equal expected progeny sex ratio, $x_0 =
0.8$, and the numbers of males and females are the same.
The lines are obtained from Monte Carlo calculation with population size $N =
10^3$, and the linepoints are from the integrodifference
equations [Eqs.~\eqref{eq:pdf1} and \eqref{eq:pdf2}]
started with $\phi^m(x,t=0) = \phi^f(x,t=0) = \delta (x-x_0)/2$.
(b) Pearson correlation coefficient of the
offspring sex ratio between parents and children, calculated from the Monte
Carlo simulation for various values of $\mu$. The population size is $N=10^3$,
and we have used $10^2$ equilibrated samples. The offspring sex ratio of a
father is positively correlated with that of his sons who have offspring, and
the degree of correlation decreases as $\mu$ grows. No such correlation exists
between mothers and daughters.}
\label{fig:mc}
\end{figure}

Although Eq.~\eqref{eq:shaw} illustrates the basic mechanism of Fisher's theory,
a more detailed view is provided by genetic
models~\cite{eshel1975selection,charnov1982theory,karlin1986theoretical},
of which we
will investigate the simplest one called a haploid model~\cite{seger2002models}.
As a Monte Carlo version of it, let us consider a population of $N$ individuals
with the following assumptions:
(i) Every individual $i$ has two attributes, i.e., one is the allele related
with the expected progeny ratio denoted by $x_i$, and the other is the sex.
(ii) For each mating event, we randomly choose a male and a female as parents.
(iii) The resulting offspring inherits either $x_\text{father}$ or
$x_\text{mother}$ equally probably, and (iv) the sex is male with probability
$x_\text{father}$.
(v) With probability $\mu \ll 1$, mutation changes $x_i$
to a random number drawn from a probability density function on the unit
interval, which we choose to be the uniform distribution $\mathcal{U}(0,1)$ for
the sake of analytic tractability.
(vi) One
generation consists of $N$ mating events to produce $N$ individuals of the
offspring generation, and an individual may be chosen to mate more than once.
This is a model of nonoverlapping generations in the sense that
the offspring generation completely replaces the parental one, which is common
in many evolutionary models.

The first three assumptions are already found in the evolutionary-stability
argument [see, e.g., Eq.~\eqref{eq:shaw}]. On the other hand, we adopt from
Ref.~\onlinecite{seger2002models} the fourth assumption that
only one parent's allele is relevant to the expected progeny sex ratio.
Yet the difference from
Ref.~\onlinecite{seger2002models} is that we regard the
father as the relevant side, which has been supported by empirical
studies~\cite{trichopoulos1967evidence,khoury1984paternal,gellatly2009trends}.
Note that this is the point where the symmetry between males and females is
broken. Most importantly, it is purely hypothetical that the expected
progeny sex ratio is determined by a single gene as in this model (see, however,
Ref.~\onlinecite{gellatly2009trends} for more discussion).
With regard to the fifth assumption, such memoryless mutation with full
variation within the unit interval would certainly be ideal, but we can always
think of an effective mutation rate with which the genetic information is lost.
We will see below that the choice of the uniform distribution greatly simplifies
our analysis in calculating the average effect of mutation.

A typical simulation result is shown in Fig.~\ref{fig:mc}(a), where one can
see the average fraction of males, denoted by $r$, approach $1/2$ as time $t$
goes by, even if the system starts from a state far from $x=1/2$.
However, if one measures the average carefully, $r(t\to \infty)$ is actually
slightly above $1/2$,
as will be detailed below. Before proceeding, we stress that
this Monte Carlo approach provides detailed information of the
population. For example, we can trace the offspring sex ratio of a father
and compare it with that of his son.
The correlation in their offspring sex ratios
can thus be calculated as a function of $\mu$ [Fig.~\ref{fig:mc}(b)]. The ratios
are positively correlated between fathers and sons, whereas they are not
between mothers and daughters, in accordance with
Ref.~\onlinecite{gellatly2009trends}.

\subsection{Integrodifference equations}

\begin{figure}
\includegraphics[width=0.49\columnwidth]{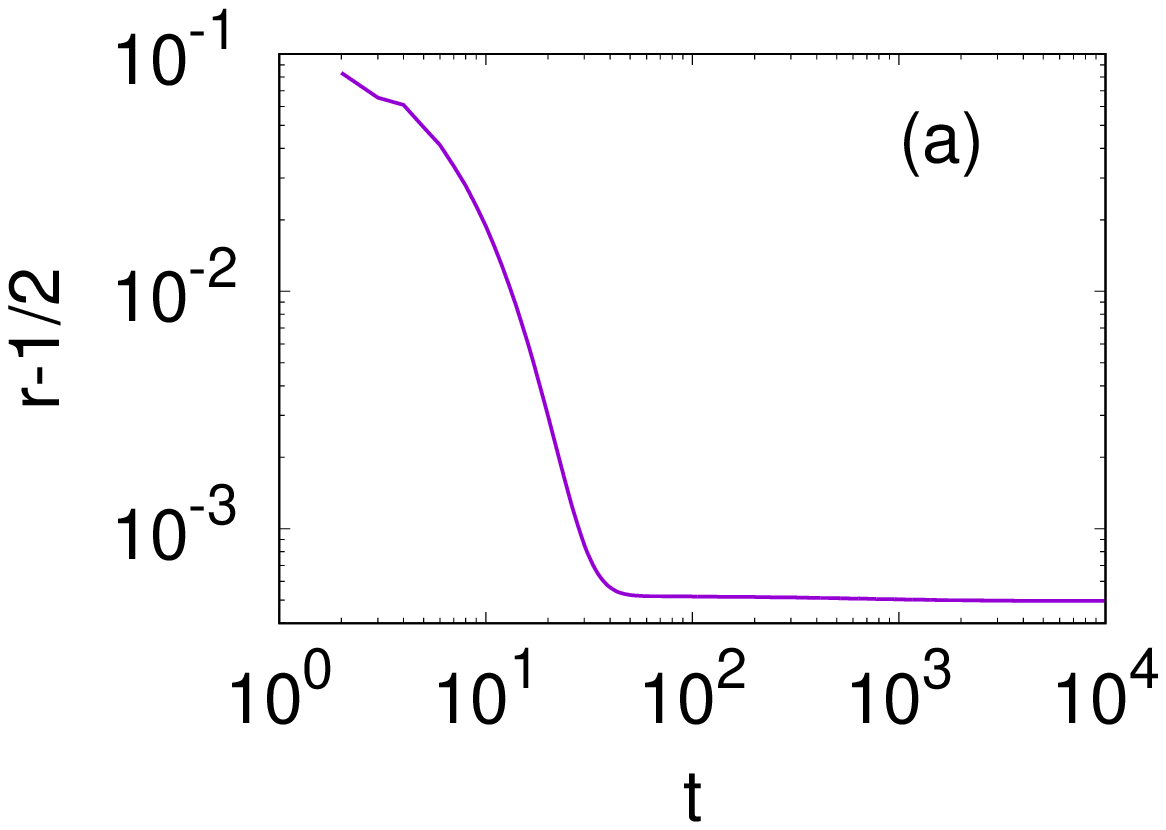}
\includegraphics[width=0.49\columnwidth]{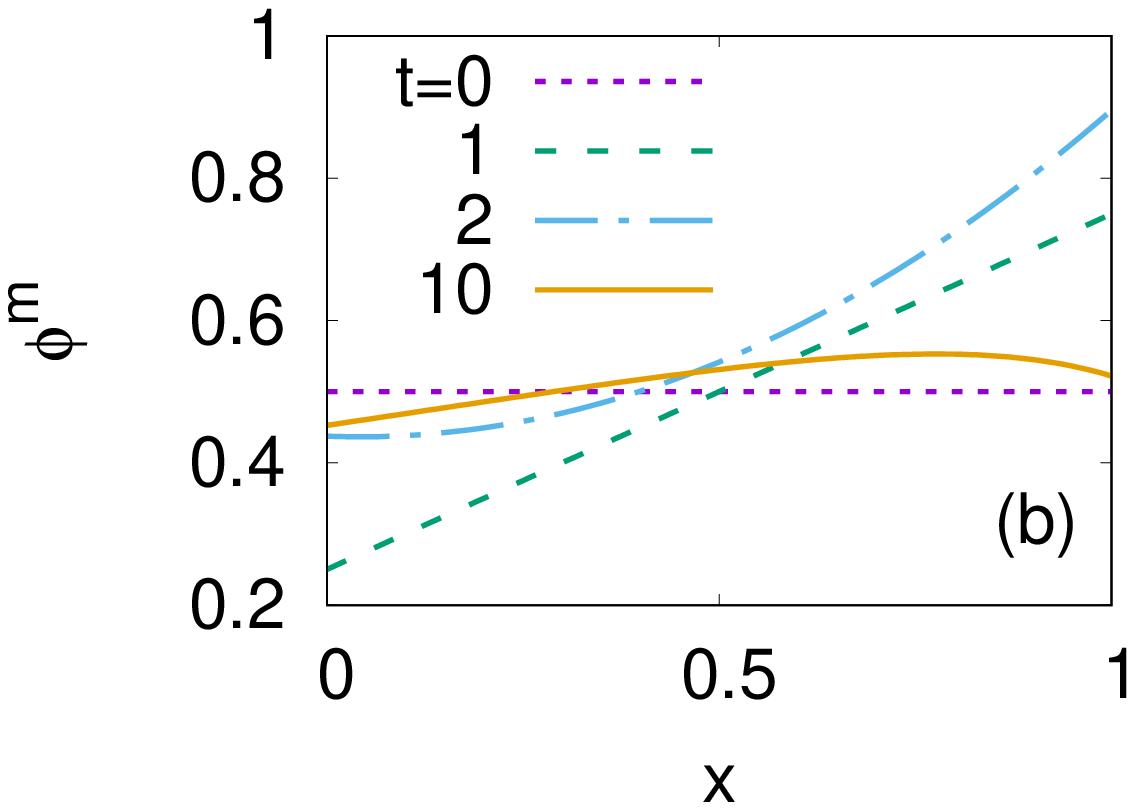}
\includegraphics[width=0.49\columnwidth]{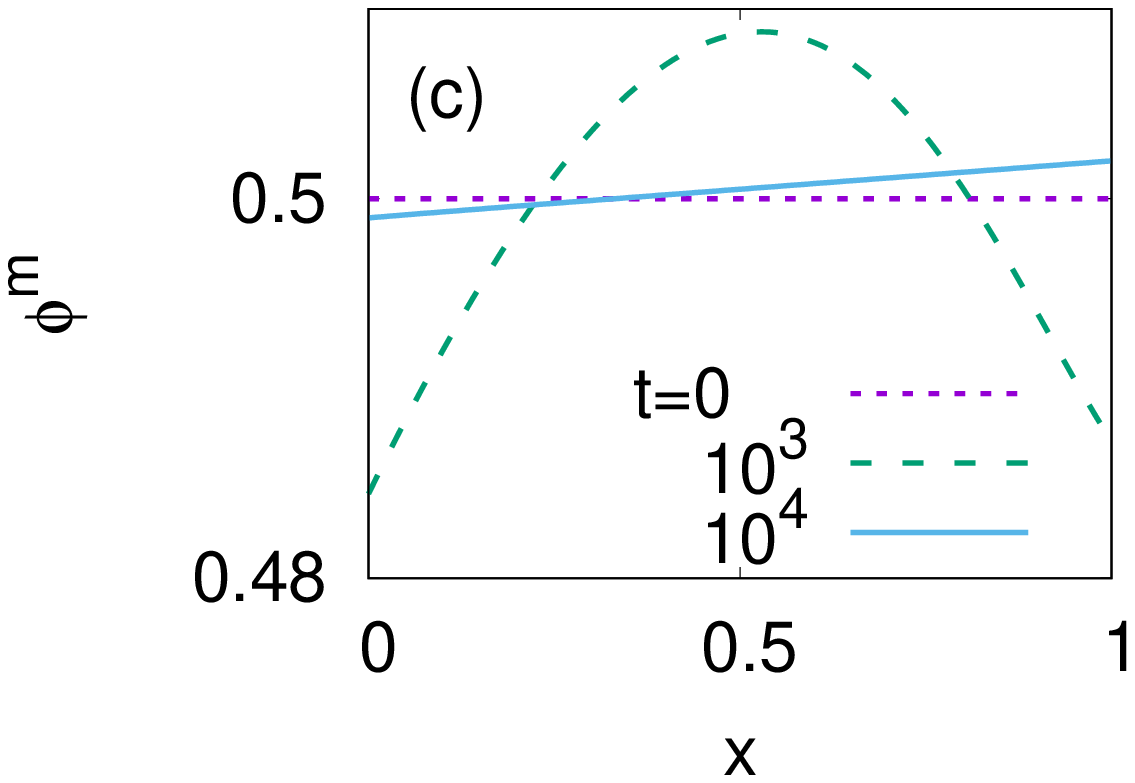}
\includegraphics[width=0.49\columnwidth]{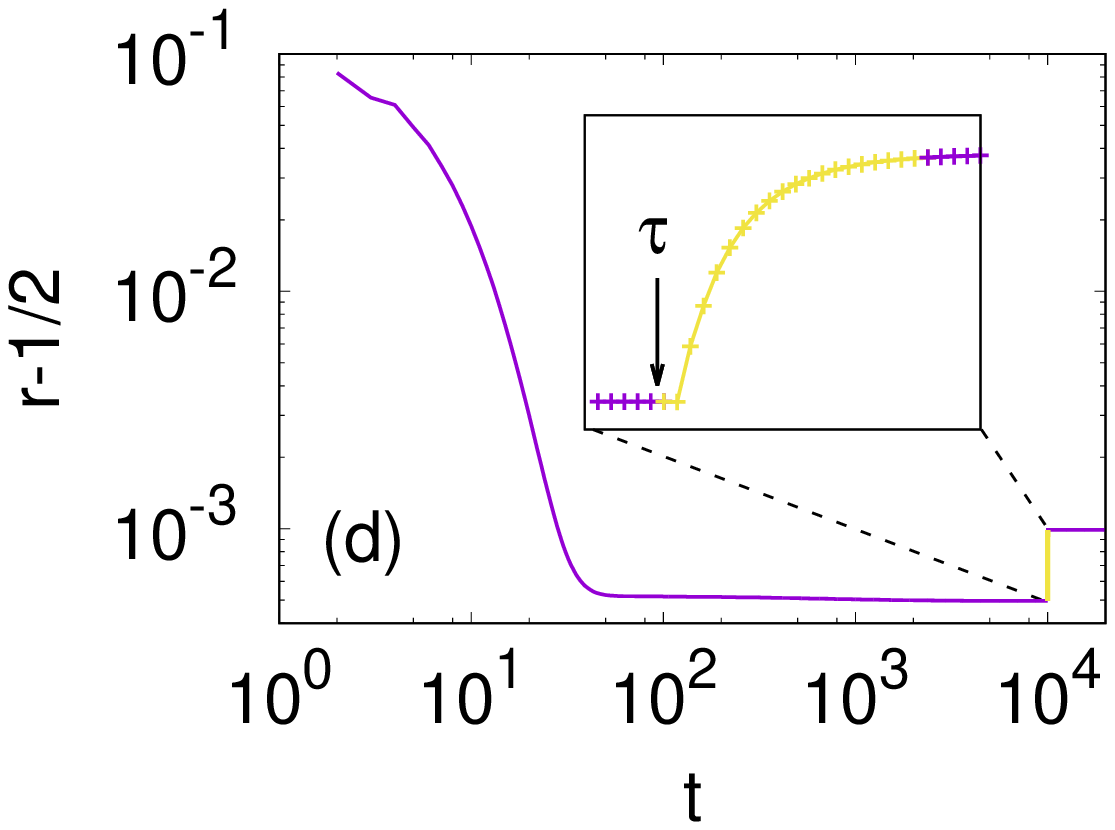}
\caption{(Color online)
(a) The sex ratio deviation from $1/2$, obtained by iterating
Eqs.~\eqref{eq:pdf1} and \eqref{eq:pdf2}. The initial condition is given as
$\phi^m(x,t=0) = \phi^f(x,t=0) = 1/2$.
Note that $r(t=1)$ is exactly $1/2$ for this initial condition
because $r(t+1) = \left< x \right>_m$.
We use the trapezoidal rule~\cite{newman2013computational}
in evaluating integrals such as $r(t)$ and $\left< x \right>_m$.
(b) Short-time and (c) long-time evolution of $\phi^m(x,t)$
from the same uniform random initial condition as in (a).
(d) Transition to another stationary state when $\mu$ changes to $\mu'=2\mu$
immediately after time $\tau=10^4$. Inset: Zoomed view around $t=\tau$, where
the first 20 generations after the change are drawn in light yellow.
}
\label{fig:numeric}
\end{figure}

To observe this deviation without statistical fluctuations, let us deal with
an infinite population. We define $\phi^m(x,t)dx$ as the
probability of being male with an expected progeny sex ratio $\in
(x, x+dx)$ at generation $t$. The fraction of males in
the total population will thus be $r(t) = \int_0^1 \phi^m(x,t) dx$.
We define $\phi^f(x,t)$ as the female counterpart, together with the fraction of
females, $\int_0^1 \phi^f (x,t) dx = 1 - r(t)$. According to the population
dynamics given above, the time evolution is described by
the following integrodifference equations in the absence of mutation:
\begin{eqnarray}
\phi_{\mu=0}^m(x,t+1) &=& \frac{1}{2} \left[
\frac{\phi^m(x,t)}{r(t)} x +
\frac{\phi^f(x,t)}{1-r(t)} \left<x\right>_m \right] \label{eq:phi0m}\\
\phi_{\mu=0}^f(x,t+1) &=& \frac{1}{2} \left[
\frac{\phi^m(x,t)}{r(t)} \tilde{x} +
\frac{\phi^f(x,t)}{1-r(t)} \left< \tilde{x} \right>_m
\right], \label{eq:phi0f}
\end{eqnarray}
where
$\left< x \right>_m \equiv \int_0^1 x \phi^m (x,t) / r(t) dx$
and $\left< \tilde{x} \right>_m \equiv 1- \left< x \right>_m$. When mutation
happens to individuals randomly drawn without replacement, the full dynamics can
be written as
\begin{eqnarray}
\phi^m(x,t+1) &=& (1-\mu) \phi_{\mu=0}^m(x,t+1) + \mu \left< x \right>_m
\label{eq:pdf1}\\
\phi^f(x,t+1) &=& (1-\mu) \phi_{\mu=0}^f(x,t+1) + \mu \left< \tilde{x} \right>_m
\label{eq:pdf2},
\end{eqnarray}
where $\mu$ is the mutation rate. The right-hand sides of
Eqs.~\eqref{eq:pdf1} and \eqref{eq:pdf2} are determined by
$\phi_{\mu=0}^m$ and $\phi_{\mu=0}^f$ at $t$, as one
can see by plugging Eqs.~\eqref{eq:phi0m} and \eqref{eq:phi0f} there.
Note that if one integrates Eq.~\eqref{eq:pdf1} over $x$, with $\phi_0^m$ given
in Eq.~\eqref{eq:phi0m}, it correctly leads to
$r(t+1) = \left< x \right>_m$,
confirming that fathers determine the progeny sex ratio.
If we start from uniform distribution $\phi^m (x,t=0) = \phi^f (x,t=0) = 1/2$,
then numerical iteration of Eqs.~\eqref{eq:pdf1} and \eqref{eq:pdf2} shows that
$r(t)$ converges to a stationary value away from $1/2$ as $t \to \infty$
[Fig.~\ref{fig:numeric}(a)].
The stationary sex ratio is well fitted by the
least-squares method to
\begin{equation}
r^{\text{fit}}_\infty (\mu) \approx 1/2 + 0.5 \mu - 2.8 \mu^2.
\label{eq:fit}
\end{equation}
It is also instructive
to look into $\phi^m(x,t)$ itself: At $t \lesssim O(10)$,
individuals with larger $x$ occupy higher portions in $\phi^m(x,t)$ because
they are more likely to produce male offspring [Fig.~\ref{fig:numeric}(b)].
This effect competes with the loss of genetic contribution in the Fisherian
mechanism, generating a unimodal shape at $t \sim O(10)$.
These two effects eventually balance each other by making
$\phi^m(x,t)$ a linear function of $x$ with a small positive slope
[Fig.~\ref{fig:numeric}(c)]. In Appendix~\ref{app:stat},
we show how one can find the functional forms of the stationary
distributions $\phi_\text{st}^m$ and $\phi_\text{st}^f$ as Taylor series.
Although it takes long from the uniform initial condition to this stationary
state, the distance between stationary states of different $\mu$'s is
relatively short [Fig.~\ref{fig:numeric}(d)].

\subsection{Renormalization analysis}

To understand the behavior in Eq.~\eqref{eq:fit},
let us assume that the mutation rate $\mu$
is so low that the population may have only two alleles at most, i.e., one is
resident and denoted by $A$, and the other is mutant and denoted by $a$.
These alleles are related to the expected progeny sex ratio but
independent of the probability for the carrier to be a parent of the next
generation~\cite{hartl1981primer}.
Let $x$ and $X$ be the expected progeny sex ratios of
$a$ and $A$, respectively. The allele $a$ is observed with frequency $q_m$
among males and with $q_f$ among females.
The possibilities of mating events are summarized in Table~\ref{tab:haploid}.
\begin{table}
\caption{Frequencies and the progeny types of the
four mating cases in the haploid model with two alleles $a$ and $A$.
A male with the mutant allele
$a$ will have a son with probability $x$, whereas the probability is
$X$ for a male with the resident allele $A$. We have defined
$\tilde{X} \equiv 1-X$ and $\tilde{x} \equiv 1-x$.}
\begin{tabular*}{\hsize}{@{\extracolsep{\fill}}cccccc}\hline\hline
 & & \multicolumn{2}{c}{daughters} &  \multicolumn{2}{c}{sons}
 \\\cline{3-6}
$\mars \times \female$       & frequency & $a$     & $A$    & $a$ & $A$\\\hline
$a \times a$ & $q_m q_f$     & $\tilde{x}$ &        & $x$ &    \\
$a \times A$ & $q_m (1-q_f)$ & $\frac{1}{2}\tilde{x}$ &
$\frac{1}{2}\tilde{x}$ &
$\frac{1}{2} x$ & $\frac{1}{2}x$ \\
$A \times a$ & $(1-q_m) q_f$ & $\frac{1}{2}\tilde{X}$ &
$\frac{1}{2}\tilde{X}$
& $\frac{1}{2} X$ & $\frac{1}{2}X$ \\
$A \times A$ & $(1-q_m)(1-q_f)$ & & $\tilde{X}$ &  & $X$ \\\hline\hline
\end{tabular*}
\label{tab:haploid}
\end{table}
Using this table,
one can calculate the frequencies of $a$ in the next generation as
follows~\cite{seger2002models}:
\begin{eqnarray}
q_m' &=& \frac{q_m q_f x + \frac{1}{2} q_m (1-q_f)x + \frac{1}{2}(1-q_m)
q_f X}{q_m x + (1-q_m)X},\label{eq:recur1}\\
q_f' &=& \frac{q_m q_f \tilde{x} + \frac{1}{2} q_m (1-q_f) \tilde{x} +
\frac{1}{2}(1-q_m) q_f \tilde{X}}{q_m \tilde{x} + (1-q_m)
\tilde{X}}\label{eq:recur2}.
\end{eqnarray}
For example, the expected fraction of male offspring is obtained from the last
two columns as
\begin{eqnarray}
r &=& q_m q_f x + q_m (1-q_f)  x\nonumber\\
&+& (1-q_m) q_f X + (1-q_f)(1-q_m) X\\
&=& q_m x + (1-q_m) X,
\label{eq:son}
\end{eqnarray}
which is the denominator of Eq.~\eqref{eq:recur1}. Likewise, the
probability to have male offspring with allele $a$ is calculated from the second
last column of Table~\ref{tab:haploid}, which is the numerator of
Eq.~\eqref{eq:recur1}.

\begin{figure}
\includegraphics[width=0.6\columnwidth]{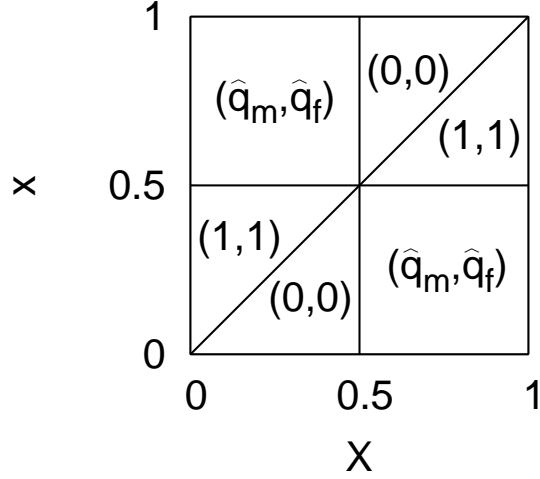}
\caption{Stable fixed points of Eqs.~\eqref{eq:recur1} and \eqref{eq:recur2}
represented on the $(X,x)$ plane.
For example, if $(X,x)=(0.4,0.3)$, then the system will converge to
$(q_m, q_f) = (0,0)$.
}
\label{fig:fp}
\end{figure}

The system of Eqs.~\eqref{eq:recur1} and \eqref{eq:recur2} has three fixed
points:
\begin{equation}
(q_m, q_f) = (0,0), (1,1), \left( \hat{q}_m, \hat{q}_f \right)
\label{eq:fp}
\end{equation}
where $\hat{q}_m \equiv (X-1/2)/(X-x)$ and $\hat{q}_f \equiv 2\tilde{x}
\hat{q}_m$.
The first fixed point is important in the context of invasion-fixation dynamics
because both $q_m$ and $q_f$ are small when $a$ is newly introduced at
$t=0$.
We thus linearize Eqs.~\eqref{eq:recur1} and \eqref{eq:recur2} in the
vicinity of $(q_m, q_f) = (0,0)$ to obtain
\begin{equation}
\begin{pmatrix}
q_m' \\ q_f'
\end{pmatrix}
=\frac{1}{2}
\begin{pmatrix}
x/X & 1\\
\tilde{x}/\tilde{X} & 1\\
\end{pmatrix}
\begin{pmatrix}
q_m \\ q_f
\end{pmatrix}.
\end{equation}
It is straightforward to obtain the eigenvalues $\lambda_{\pm}$
with $\lambda_+ \ge \lambda_-$ and the corresponding eigenvectors.
The instability threshold of the fixed point $(q_m, q_f) = (0,0)$ is
characterized by $\lambda_+ = 1$. In this case, a little algebra shows
\begin{equation}
1 = \frac{1}{2} \left( \frac{x}{X} + \frac{\tilde{x}}{\tilde{X}} \right),
\label{eq:shaw2}
\end{equation}
which is equivalent to Eq.~\eqref{eq:shaw} with $C_1 =
C_2$~\cite{seger2002models}.
If $X = 1/2 + \epsilon$ with $|\epsilon| \ll 1$, then the eigenvalues are
approximated to the first order of $\epsilon$ as
\begin{eqnarray}
\lambda_+ &\approx& 1 + 2\left(\frac{1-2x}{1+2\tilde{x}} \right) \epsilon\\
\lambda_- &\approx& \left(x-\frac{1}{2} \right) - 2 \tilde{x} \left(
\frac{1+2x}{1+2\tilde{x}} \right) \epsilon,
\end{eqnarray}
and the eigenvectors are
\begin{eqnarray}
\vec{v}_+ &\approx& \left( 1, 2\tilde{x} + 4\tilde{x}
\left( \frac{1+2x}{1+2\tilde{x}} \right)\epsilon \right)\\
\vec{v}_- &\approx& \left( 1, -1-4 \left( \frac{1-2x}{1+2\tilde{x}} \right)
\epsilon \right).
\end{eqnarray}
If any of $\lambda_\pm$ exceeds one, then the mutant can invade the population.
It happens either when $X>1/2$ and $x<X$, or when $X<1/2$ and $x>X$, which
implies that the sex ratio tends to $1:1$ in agreement with Eq.~\eqref{eq:shaw}.
The linear-stability analysis can be applied to the other fixed points as well,
whereby we conclude that the relevant fixed point is $(q_m, q_f) = (0,0)$ or
something close to it, as far as $\epsilon$ is sufficiently small
(Fig.~\ref{fig:fp}).

\begin{figure}
\includegraphics[width=0.49\columnwidth]{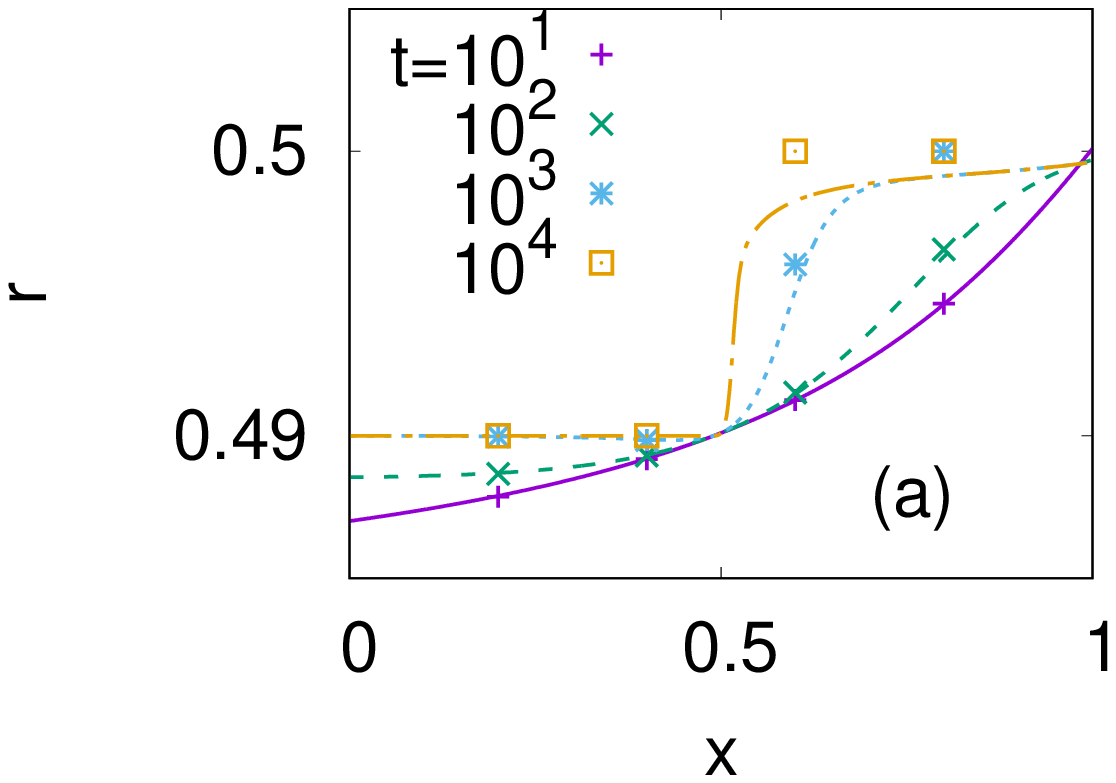}
\includegraphics[width=0.49\columnwidth]{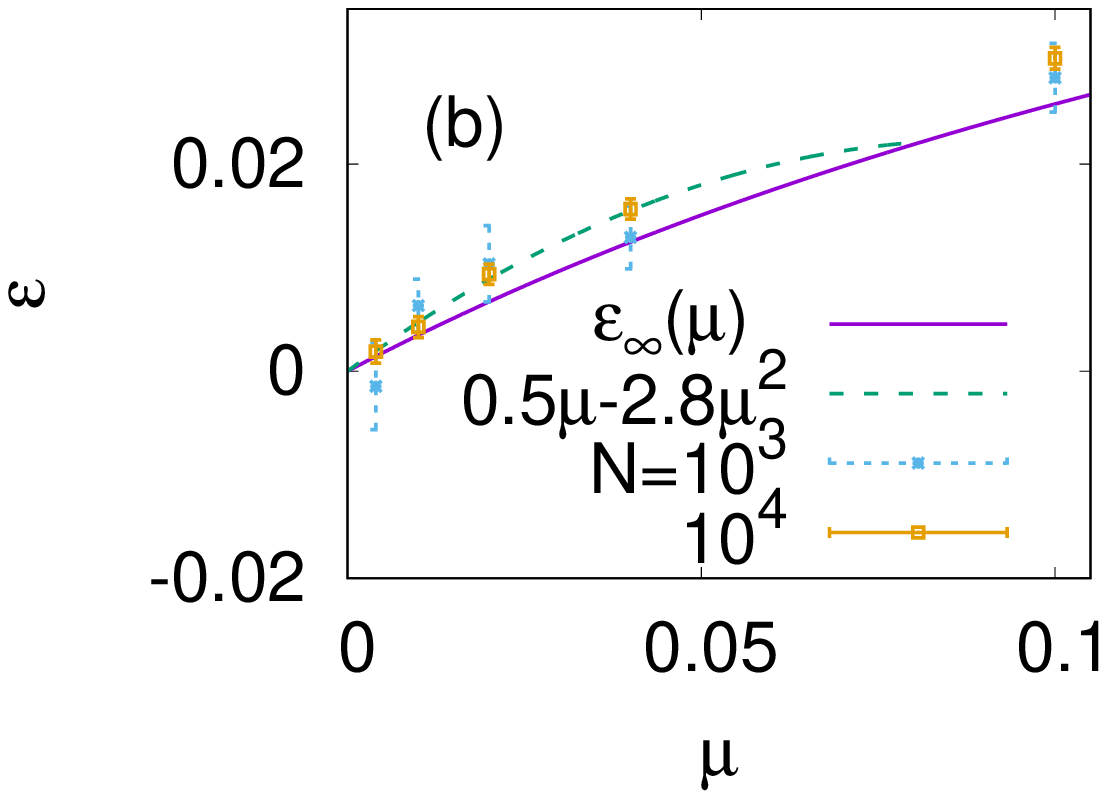}
\caption{(Color online)
(a) Time evolution of the fraction of males [Eq.~\eqref{eq:son}]
from the direct recursion [Eq.~\eqref{eq:recur1} and \eqref{eq:recur2}] (points)
and its approximation [Eq.~\eqref{eq:qm}] (lines),
when $q_f^{(t=t_0)} = q_m^{(t=t_0)} = 0.01$ and $X=0.49$ in the haploid model
with two alleles.
(b) Deviation from the Fisherian ratio as a function of $\mu$.
The solid line shows the approximation in Eq.~\eqref{eq:noq}. For
comparison, we also plot $0.5\mu - 2.8\mu^2$ of Eq.~\eqref{eq:fit}.
The points with error bars show results from the
Monte Carlo version of the haploid model with population size $N$.
The errors are estimated over $50$ equilibrated samples.
}
\label{fig:r}
\end{figure}

When a small number of mutants have appeared, $(q_m, q_f)$ will be aligned
along $\vec{v}_+ \approx (1,2\tilde{x})$ by the fast dynamics with a timescale
of $t_0 \equiv \left| \ln \lambda_- \right|^{-1} \sim O(1)$.
Because $\vec{v}_- \approx
(1,-1)$, the sum of $q_m$ and $q_f$ is approximately conserved during this
alignment, after which
\begin{equation}
\left( q_m^{(t=t_0)}, q_f^{(t=t_0)} \right) \approx Q\left(
\frac{1}{1+2\tilde{x}}, \frac{2\tilde{x}}{1+2\tilde{x}}\right),
\label{eq:init}
\end{equation}
where $Q \equiv q_m^{(t=0)} + q_f^{(t=0)}$
is the initial fraction of mutants.
The system then slowly approaches the fixed point $(0,0)$, which means that the
mutants go extinct. From the fact
that $\lambda_+ \approx 1$, we see that the characteristic timescale diverges
in this slow dynamics.
To be more precise, the trajectory can be expressed by
\begin{equation}
q_f \approx \left[2\tilde{x} +4\tilde{x} \left(
\frac{1+2x}{1+2\tilde{x}} \right) \epsilon \right] q_m + C q_m^2,
\label{eq:quad}
\end{equation}
with
\begin{eqnarray}
C &\approx& -\frac{2\tilde{x}(1-2x)(1+2x)}{1+2\tilde{x}}\nonumber\\
&-&
\frac{4\tilde{x} (21-78x+180x^2-168x^3+32x^4)}{(1+2\tilde{x})^3} \epsilon
\label{eq:C}
\end{eqnarray}
to the order of $\epsilon$. Note that we have to keep the order of $q_m^2$.
Plugging Eqs.~\eqref{eq:quad} and \eqref{eq:C} into
Eq.~\eqref{eq:recur1} and using the continuous-time approximation,
we get the following differential equation:
\begin{equation}
\frac{dq_m}{dt} \approx c_1 q_m + c_2 q_m^2,
\label{eq:ode}
\end{equation}
where
\begin{eqnarray}
c_1 &\equiv& 2\left(\frac{1-2x}{1+2\tilde{x}}\right) \epsilon \label{eq:c1}\\
c_2 &\equiv& -\frac{(1-2x)^2}{1+2\tilde{x}}
-\frac{6(7-10x+12x^2-8x^3)}{(1-2x)^{-1}(1+2\tilde{x})^3} \epsilon.
\label{eq:c2}
\end{eqnarray}
One can readily solve Eq.~\eqref{eq:ode} to find
\begin{equation}
q_m(t) = \frac{c_1 e^{c_1 t} q_m^{(t=t_0)}}{c_1 e^{-c_1 t_0} - c_2 (e^{c_1
t} - e^{-c_1 t_0}) q_m^{(t=t_0)}}.
\label{eq:qm}
\end{equation}
In the limit of $\epsilon \to 0$,
the timescale of this dynamics diverges because $q_m(t) \sim t^{-1}$.
In addition, if $c_1>0$, Eq.~\eqref{eq:qm} converges to
\begin{equation}
\lim_{t \to \infty} q_m (t) = -c_1/c_2 \approx 2\epsilon / (1-2x),
\label{eq:limit}
\end{equation}
which coincides with the correct result, $\hat{q}_m$ in
Eq.~\eqref{eq:fp}, to the order of $\epsilon$.
Now, we have an approximate expression for the male fraction as a function
of time by substituting Eq.~\eqref{eq:qm} into Eq.~\eqref{eq:son}. It may be
written as $r(t|X,x)$ to emphasize that it is also conditioned by $X$ and $x$.
Although this result involves uncontrolled approximations such as
Eq.~\eqref{eq:init} and $t_0 \approx 0$,
the formula works reasonably well as shown in Fig.~\ref{fig:r}(a).

Now imagine that the population initially had $X = X_0$.
Random mutation occurs with a timescale $t \sim O(\mu^{-1})$ at any point of the
population, and the sex ratio
will be renormalized as a response to mutation as follows:
\begin{eqnarray}
\epsilon_{k+1} &=& \int_0^1 r \left(t=\mu^{-1} | X=X_k, x\right) dx -
\frac{1}{2}\\
&=& \int_0^1 \left[ \epsilon_k + (x-X_k) q_m \left(t= \mu^{-1}|X_k,x \right)
\right] dx\label{eq:integral}\\
&\equiv& E(\epsilon_k,\mu),
\label{eq:iter}
\end{eqnarray}
where we have defined $\epsilon_k
\equiv X_k - 1/2$ with an integer index $k=0,1,\ldots$.
As an example, assume that
$\mu$ can be made arbitrarily small to satisfy $\mu \ll |c_1|$ all the time.
According to the approximate expression given above, as $t \to \infty$,
the male fraction $r$ converges to $\epsilon_k + 1/2$ when $c_1 > 0$, and
to $1/2 + O(\epsilon_k^2)$ otherwise [see, e.g.,
Fig.~\ref{fig:r}(a)]. As a result, we have approximately
$\frac{1}{2} \left[ \left( \epsilon_k +
\frac{1}{2} \right) + \frac{1}{2} \right]$ on the right-hand side of
Eq.~\eqref{eq:iter}, which is to be identified with
$X_{k+1} = 1/2 + \epsilon_{k+1}$.
The map obviously flows into $\epsilon_\infty = 0$, and we thus
conclude that the system achieves the Fisherian ratio $X=1/2$ in this
limit of $\mu \to 0$.
Having observed this limiting case, we assume that the right-hand side of
Eq.~\eqref{eq:iter} can still be approximated by a linear function of
$\epsilon_k$ for finite $\mu$, i.e.,
\begin{equation}
E(\epsilon_k, \mu) \approx U(\mu) \epsilon_k + V(\mu)
\label{eq:linear}
\end{equation}
when $\epsilon_k \ll 1$. If this assumption holds, then we have
\begin{equation}
\epsilon_k = U^k (\mu) \epsilon_0 + \sum_{l=0}^{k-1} U^l(\mu) V(\mu),
\end{equation}
and the ``dressed'' value converges to
\begin{equation}
\epsilon_\infty (\mu) = \frac{V (\mu)}{1-U (\mu)}
\label{eq:converged}
\end{equation}
as long as $\left| U(\mu) \right| < 1$.
From Eq.~\eqref{eq:linear}, we may write
$V(\mu) = \lim_{\epsilon_k \to 0} E(\epsilon_k,\mu)$ and
$U(\mu) = \left. \partial E/\partial \epsilon_k
\right|_{\epsilon_k=0}$,
both of which give closed-form expressions if the integral and the limiting
process of $\epsilon_k \to 0$ commute with each other (see
Appendix~\ref{app:uv} for details). Then, we
take another limit of $Q \to 0$ and get our main result,
\begin{equation}
\epsilon_\infty (\mu) \approx \frac{3\mu(\ln 3-1)}{14- 12\ln 3 + 3\mu \ln 3},
\label{eq:noq}
\end{equation}
which gives $\epsilon_\infty (\mu) \approx 0.36 \mu - 1.46 \mu^2$
for $\mu \ll 1$.
Note the order of the limiting processes: If we had taken this zero-$Q$ limit
from the beginning, then the result would have been trivially zero.
In Fig.~\ref{fig:r}(b), we see that Eq.~\eqref{eq:noq} correctly
captures the qualitative behavior of the Monte Carlo results.

\section{Discussion}
\label{sec:discuss}

We have investigated a model designed to support Fisher's theory, and it turns
out that a small correction $\propto \mu$ has to be added.
The reason for this correction is that the system
reacts differently to female-biasing and male-biasing mutants, as already
implied in Eq.~\eqref{eq:init}: When $X\approx 1/2$,
if we compare female-biasing mutants, say, with $x = 0.4$, and
male-biasing ones with $x=0.6$, then $q_m$ will be greater in the latter case.
On average, therefore, the male fraction is likely to be experienced
as greater than $1/2$.
Recall that the asymmetric part of the model is the father's predominance in
determining the offspring's sex. We have shown that the system nevertheless
becomes symmetric in a limit of $\mu \to 0$, which is the message of the
fixed-point analysis in Eq.~\eqref{eq:shaw2}.
In this sense, Fisher's theory can be thought of as a symmetry
preservation mechanism.
At the same time, each individual has an internal variable,
the expected progeny ratio $x$. An interesting point is that this internal
variable experiences little selection pressure when the sex ratio is 1:1,
so that the gene pool can retain a high degree of genetic diversity [see, e.g.,
Fig.~\ref{fig:numeric}(c)].

For many species with female-biased sex ratios, the bias has been
successfully explained within the Darwinian framework, e.g., by local mate
competition~\cite{west2009sex}.
On the other hand,
the human sex ratio is slightly biased toward males~\cite{wdb},
which is also believed to have an evolutionary origin.
We have already seen how the Fisherian mechanism maintains an (almost) equal
sex ratio at birth. If we furthermore assume that
males have a higher mortality rate than females in their
youth~\cite{stinson1985sex}, then
Fisher's equal-investment theory predicts a male-biased sex ratio at
birth: Otherwise, the overall investment in male offspring would eventually
become smaller than in female ones~\cite{west2009sex}.
Unfortunately, empirical verification of this prediction is exceedingly
complicated by the difficulty of measuring parental
investment~\cite{sieff1990explaining}.
This work has proposed another mechanism that induces a male-biased sex
ratio.
In case of diploid organisms like humans, the proportionality coefficient in
front of $\mu$ will
depend on the dominance between the resident and mutant alleles, but it is
plausible that our estimate from the haploid model sets an upper bound for it
because a recessive mutant would not much perturb the system.
If we naively guess that our mechanism is responsible for the
commonly known human sex ratio $\approx 1.07:1$ at birth, then the effective
mutation rate will be $\mu \gtrsim 0.05$, meaning that the
allele of the expected progeny sex ratio will be mutated roughly in 20
generations.
We also note that one can empirically measure
the correlation of offspring sex ratios in families,
as we have depicted in Fig.~\ref{fig:mc}(b).
After suitable modification of the modeling assumptions, this sort of
Monte Carlo calculation may be compared with genealogical data to estimate
$\mu$.
One can also monitor how the sex ratio varies when mutations are induced by
chemicals or radiation. For example, the human mutation rate showed a twofold
increase among individuals involved in the Chernobyl accident even at a
conservative
estimate~\cite{dubrova1996human,*weinberg2001very,*moller2006biological}. A
recent investigation demonstrates that the sex ratio increased after the
accident~\cite{scherb2013increased}, which seems consistent with our study.

In a more general context, our study
suggests that the conventional fixed-point analysis, focusing on
a static equilibrium, may not catch the exact picture if perturbative
effects are not taken into account, and that the behavior can be
explained by renormalizing the fluctuations around the fixed point.
Our result can also be regarded as an example of mutation-selection
balance~\cite{bertels2017discovering}, in which selection drives the system to
the fixed point while at the same time it is prevented by mutation from reaching
it. Although the mutation rate is very small, its effect is of an observable
magnitude because the approach to the fixed point has a diverging timescale.

\section{Summary}
\label{sec:summary}

To summarize, we have presented a detailed analysis of the haploid model, a
microscopic foundation of Fisher's theory of equal investment:
Although the invasion-fixation dynamics of the
haploid model explains the $1:1$ ratio in the limit of $\mu \to
0$, the system reaches a
dynamic equilibrium away from the Fisherian ratio as long as $\mu$ is nonzero.
We have demonstrated this mutation-induced bias with three different approaches,
i.e., Monte Carlo simulation, integrodifference equations, and renormalization
analysis. All of these approaches give consistent results, revealing
nontrivial dynamical aspects of the Fisherian mechanism. By linking the mutation
rate and sex-ratio bias, this picture yields testable predictions, whereby
the size of this effect can be assessed empirically.

\appendix
\section{Stationary solution of the integrodifference equations}
\label{app:stat}

\begin{figure}
\includegraphics[width=0.45\columnwidth]{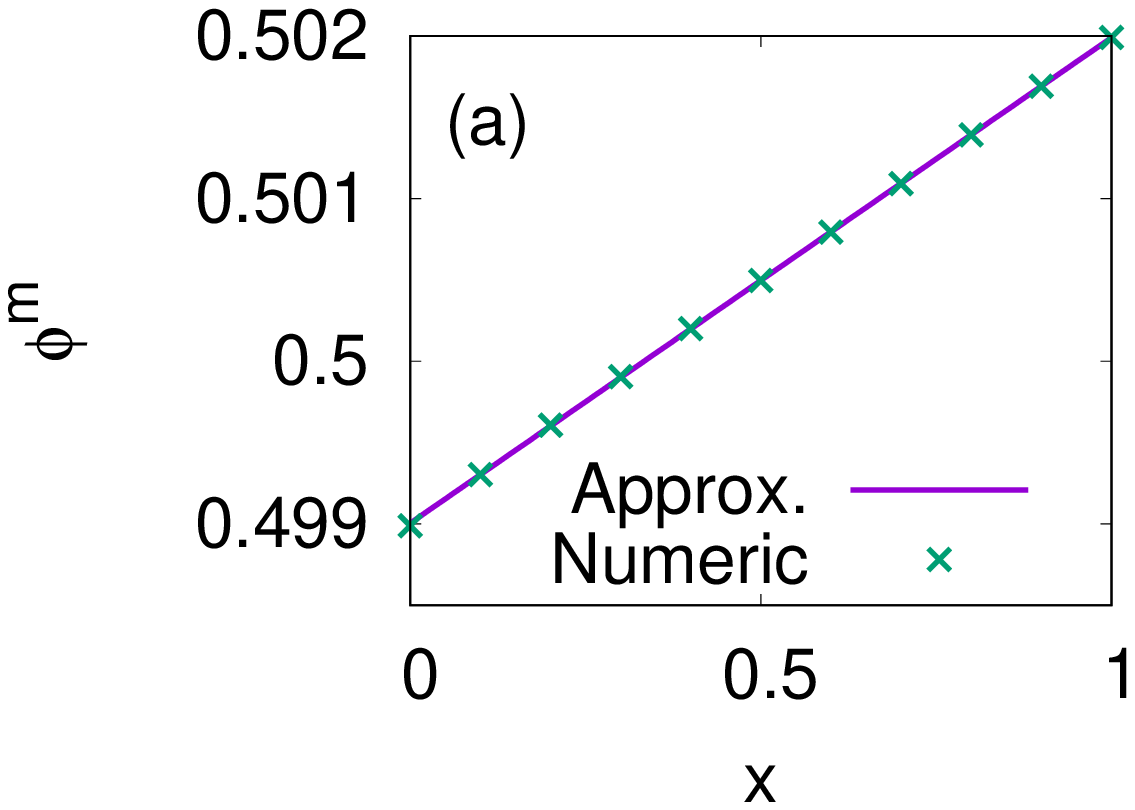}
\includegraphics[width=0.45\columnwidth]{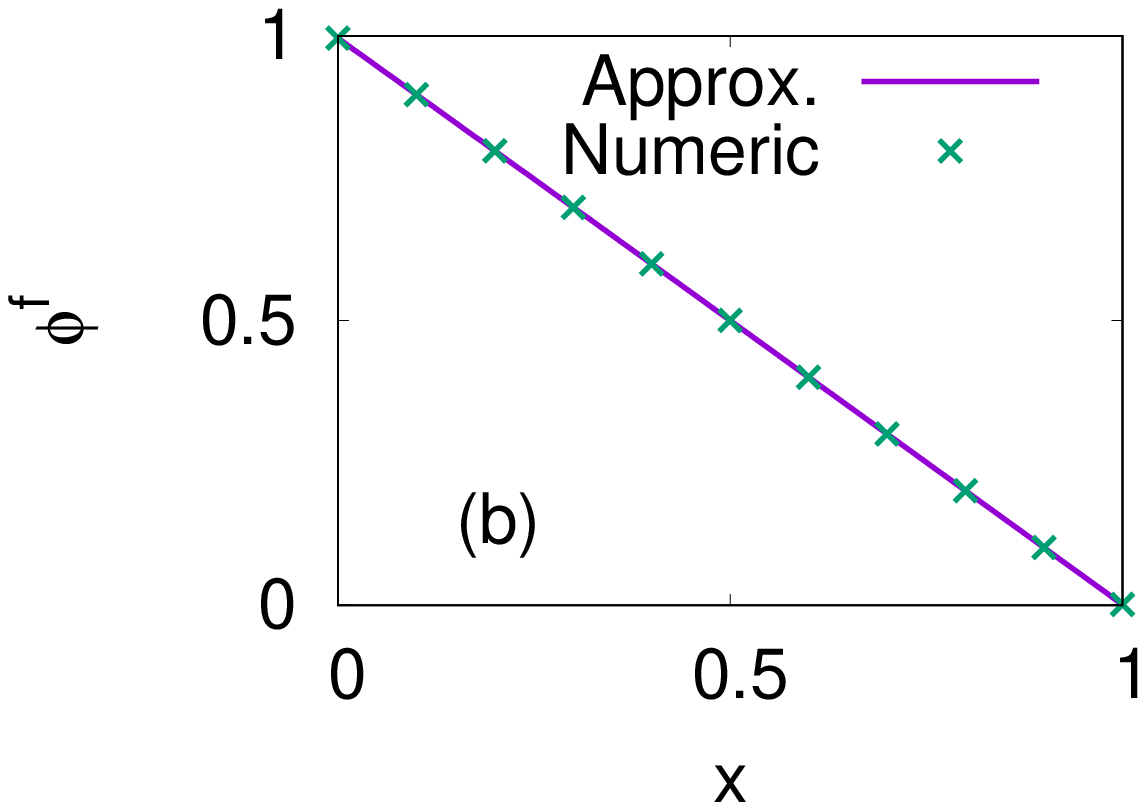}
\caption{(Color online) (a) Stationary probability density
of being males with expected progeny sex ratio $x$, and (b) that of being
females.
The points are obtained by numerical iteration of Eqs.~\eqref{eq:pdf1}
and \eqref{eq:pdf2} with the same parameters as in Fig.~\ref{fig:numeric}(a),
and the lines are drawn with the parameters that minimize Eq.~\eqref{eq:W}.}
\label{fig:lin}
\end{figure}

Let us expand the stationary distributions $\phi^m_\text{st}(x) \equiv
\phi^m(x,t\to\infty)$ and $\phi^f_\text{st}(x) \equiv \phi^f(x,t\to\infty)$
to the quadratic order:
\begin{eqnarray}
\phi^m_\text{st} (x) &\approx \alpha_0 + \alpha_1 x + \alpha_2
x^2\label{eq:lin1}\\
\phi^f_\text{st} (x) &\approx \beta_0 + \beta_1 x + \beta_2 x^2\label{eq:lin2}.
\end{eqnarray}
Within this approximation,
we actually have $5$ degrees of freedom because of the following constraint:
\begin{equation}
\int_0^1 \left[ \phi^m_\text{st} (x) + \phi^f_\text{st} (x) \right] dx =
\alpha_0  + \frac{1}{2}\alpha_1 + \frac{1}{3}\alpha_2 + \beta_0
+ \frac{1}{2}\beta_1 + \frac{1}{3}\beta_2 = 1.
\end{equation}
Let us plug Eqs.~\eqref{eq:lin1} and \eqref{eq:lin2} into
Eqs.~\eqref{eq:pdf1} and \eqref{eq:pdf2}. We define
\begin{eqnarray}
S(x) &\equiv& \alpha_0 + \alpha_1 x + \alpha_2 x^2
- \left\{ \mu
\left(
\frac{ \frac{\alpha_0}{2} + \frac{\alpha_1}{3} +
\frac{\alpha_2}{4}}{ \alpha_0 + \frac{\alpha_1}{2} + \frac{\alpha_2}{3}} \right)
+ \frac{1}{2} (1-\mu)
\left[ \frac{x (\alpha_0 + \alpha_1 x + \alpha_2 x^2)}{\alpha_0 +
\frac{\alpha_1}{2} + \frac{\alpha_2}{3}} \right. \right. \nonumber\\
&+& \left.\left.
\left(
\frac{ \frac{\alpha_0}{2} + \frac{\alpha_1}{3} +
\frac{\alpha_2}{4}}{ \alpha_0 + \frac{\alpha_1}{2} + \frac{\alpha_2}{3}} \right)
\left( \frac{\beta_0 + \beta_1 x + \beta_2 x^2}{1-\alpha_0
-\frac{\alpha_1}{2} -\frac{\alpha_2}{3} } \right)
\right]
\right\}\\
T(x) &\equiv& \beta_0 + \beta_1 x + \beta_2 x^2
- \left\{ \mu
\left( 1-
\frac{ \frac{\alpha_0}{2} + \frac{\alpha_1}{3} +
\frac{\alpha_2}{4}}{ \alpha_0 + \frac{\alpha_1}{2} + \frac{\alpha_2}{3}} \right)
+ \frac{1}{2} (1-\mu)
\left[ \frac{(1-x) (\alpha_0 + \alpha_1 x + \alpha_2 x^2)}{\alpha_0 +
\frac{\alpha_1}{2} + \frac{\alpha_2}{3}} \right. \right. \nonumber\\
&+& \left.\left.
\left( 1-
\frac{ \frac{\alpha_0}{2} + \frac{\alpha_1}{3} +
\frac{\alpha_2}{4}}{ \alpha_0 + \frac{\alpha_1}{2} + \frac{\alpha_2}{3}} \right)
\left( \frac{\beta_0 + \beta_1 x + \beta_2 x^2}{1-\alpha_0
-\frac{\alpha_1}{2} -\frac{\alpha_2}{3} } \right)
\right]
\right\}.
\end{eqnarray}
Equations~\eqref{eq:pdf1} and \eqref{eq:pdf2} mean that $S(x) = T(x) = 0$,
which will be only approximately true because Eqs.~\eqref{eq:lin1} and
\eqref{eq:lin2} are not exact. We instead minimize
\begin{equation}
W \equiv \int_0^1 dx \left[ S^2(x) + T^2(x) \right]
\label{eq:W}
\end{equation}
with respect to $\alpha_0, \alpha_1, \alpha_2, \beta_1$ and
$\beta_2$. When $\mu = 10^{-3}$, the minimum $W_\text{min} = 3.46448 \times
10^{-16}$ is found at
$\alpha_0 = 0.499004$,
$\alpha_1 = 0.00298505$,
$\alpha_2 = 2.76624 \times 10^{-10}$,
$\beta_1 = -0.989074$, and
$\beta_2 = -0.0059523$~\cite{Mathematica},
which indeed describe the stationary
solution with high precision (Fig.~\ref{fig:lin}).

\section{Evaluation of $U(\mu)$ and $V(\mu)$}
\label{app:uv}

Let us express Eq.~\eqref{eq:iter} as an integral by plugging Eq.~\eqref{eq:qm}
into Eq.~\eqref{eq:integral}. Then, we introduce $U(\mu)$ and $V(\mu)$ as
in Eq.~\eqref{eq:linear}, which implies that
$V(\mu) = \lim_{\epsilon_k \to 0} E(\epsilon_k,\mu)$ and
$U(\mu) = \left. \partial E/\partial \epsilon_k
\right|_{\epsilon_k=0}$. Provided that the integral and the limiting process of
$\epsilon_k \to 0$ commute, we can find their closed-form expressions as
follows~\cite{Mathematica}:

\begin{align}
V(\mu) &= E(0,\mu) \\
&= -\int_0^1\frac{\mu Q(-3+2x)(-1+2x)}{2[Q(1-2x)^2+\mu(3-2x)^2]} \\
&= -\frac{\mu Q}{4(\mu+Q)} \notag \\
&\times\biggl[-2(\mu+Q)+4\sqrt{\mu Q}\left\{\arctan\left(\frac{3\mu+Q}{2\sqrt{\mu Q}}\right) - \arctan\left(\frac{\mu-Q}{2\sqrt{\mu Q}}\right)\right\} \notag \\
&\quad\quad -(\mu+Q)\biggl(\ln(\mu+Q)+\ln(9\mu+Q)\biggr)\biggr],
\end{align}

\begin{align}
U(\mu)&=\frac{\partial E}{\partial \epsilon_k}\bigg|_{\epsilon_k=0} \\
&= \int_0^1\frac{2\mu^2(3-2x)^4(Q+2x-3)}{2(2x-3)\left(\mu(3-2x)^2+Q(1-2x)^2\right)^2}dx \notag \\
&\quad+\int_0^1\frac{2\mu Q(1-2x)^2\left(2Q(12x^3-16x^2+9x-6)+(2x-3)^3\right)}{2(2x-3)\left(\mu(3-2x)^2+Q(1-2x)^2\right)^2}dx \notag \\
&\quad+\int_0^1\frac{Q^2(2x-3)(1-2x)^3}{2(2x-3)\left(\mu(3-2x)^2+Q(1-2x)^2\right)^2}dx \\
&= \frac{1}{4(\mu+Q)^3} \notag \\
&\times\Bigg[\frac{2\bigl(\mu^4(18-66Q)+4\mu^3(8-9Q)Q+14\mu Q^3+Q^4+3\mu^2Q^2(9+10Q)\bigr)}{9\mu+Q} \notag \\
&\qquad - \sqrt{\mu Q}\bigl(15\mu^3+Q^2(1+Q)+\mu Q(6+5Q)+\mu^2(-3+67Q)\bigr)\notag \\
&\quad\quad\quad \times \bigg\{\arctan\biggl(\frac{\mu-Q}{2\sqrt{\mu Q}}\bigr) - \arctan\biggl(\frac{3\mu+Q}{2\sqrt{\mu Q}}\biggr)\bigg\} \notag \\
&\qquad -\mu\biggl(12(mu+Q)^3\ln3+\bigl(6\mu^3+17\mu^2 Q+2\mu(2-7Q)Q-Q^3\bigr) \notag \\
&\quad\qquad\qquad\qquad\qquad\qquad\qquad
\times\bigl(\ln(\mu+Q)-\ln(9\mu+Q)\bigr)\biggr)\Bigg].
\end{align}

We then combine these formulas as in Eq.~\eqref{eq:converged}
and take another limit of $Q\to 0$ to obtain Eq.~\eqref{eq:noq}.

\begin{acknowledgments}
H.C.J. was supported by Basic Science Research Program through the National
Research Foundation of Korea (NRF) funded by the Ministry of Education
(Grant No. NRF-2015R1D1A1A01058317).
S.K.B. was supported by Basic Science Research Program through the
National Research Foundation of Korea (NRF) funded by the Ministry of Science,
ICT and Future Planning (Grant No. NRF-2017R1A1A1A05001482).
\end{acknowledgments}

%
\end{document}